\documentclass{osa-article}
\usepackage{braket}
\usepackage{cite}
\usepackage{epstopdf}
\usepackage{amsmath}    
\usepackage{subfigure}  
\usepackage{osajournal}

\articletype{Research Article}

\begin{document}

\title{Fundamental figures of merit for engineering F\"{o}rster resonance energy transfer}
\author{Cristian L. Cortes and Zubin Jacob}

\address{Birck Nanotechnology Center, School of Electrical and Computer Engineering, Purdue University, West Lafayette, Indiana 47907, USA}

\email{\authormark{*}zjacob@purdue.edu} 



\begin{abstract}
Over the past 15 years there has been an ongoing debate regarding the influence of the photonic environment on F\"orster resonance energy transfer (FRET). Disparate results corresponding to enhancement, suppression and null effect of the photonic environment have led to a lack of consensus between the traditional theory of FRET and experiments.  Here we show that the quantum electrodynamic theory (QED) of FRET near an engineered nanophotonic environment is exactly equivalent to an effective near-field model describing electrostatic dipole-dipole interactions. This leads to an intuitive and rigorously exact description of FRET, previously unavailable,  bridging the gap between experimental observations and theoretical interpretations. Furthermore, we show that the widely used concept of Purcell factor variation is only important for understanding spontaneous emission and is an incorrect figure of merit (FOM) for analyzing FRET. To this end, we analyze the figures of merit which characterize FRET in a photonic environment 1) the FRET rate enhancement factor ($F_{ET}$), 2) FRET efficiency enhancement factor ($F_{eff}$) and 3) Two-point spectral density ($S_{EE}$) which is the photonic property of the environment governing FRET analogous to the local density of states that controls spontaneous emission. Counterintuitive to existing knowledge, we show that suppression of the Purcell factor is in fact necessary for enhancing the efficiency of the FRET  process. We place fundamental bounds on the FRET figures of merit  arising from material absorption in the photonic environment as well as key properties of emitters including intrinsic quantum efficiencies and orientational dependence. Finally, we use our approach to conclusively explain multiple recent experiments and  predict regimes where the FRET rate is expected to be enhanced, suppressed or remain the same.  Our work paves for a complete theory of FRET with predictive power for designing the ideal photonic environment to control FRET. 
\end{abstract}



\bibliographystyle{osajnl}
\bibliography{Theory2,Experiments}

\section{Introduction}
F\"orster resonance energy transfer (FRET) is the result of a near-field (1-10 nm) dipole-dipole interaction between two atoms or molecules. The outcome is an irreversible and incoherent transfer of energy from an excited-state donor atom to a ground-state acceptor atom, which manifests itself in the quenching of the donor's excited state lifetime. This phenomenon was first explained classically in 1946 \cite{forster1946energiewanderung} and then quantum mechanically in 1948 \cite{forster1948zwischenmolekulare} by T. F\"orster. Since then FRET has been the subject of intense research across all sciences; for example, in chemistry for molecular sensing; in biology for bio-imaging and the study of cellular and genetic dynamics; and in engineering for enhancing the efficiency of photovoltaic devices \cite{weissleder2008imaging,wallrabe2005imaging,bastiaens1999fluorescence,hardin2012renaissance}. 


While the microscopic nature of FRET is widely understood and accepted in simple homogeneous systems, the fundamental nature of FRET in complex photonic environments remains poorly understood and has been widely debated over the past 15 years (see Table 1).  The debate is largely based on the vastly different and seemingly contradictory results of many experiments (see Table 1) when donor-acceptor pairs are placed in the vicinity of photonic cavities or nanoparticles. In some cases FRET has been shown to be enhanced \cite{hopmeier_enhanced_1999,andrew_forster_2000,finlayson_enhanced_2001,lakowicz_radiative_2002,fujiwara_enhancement_2005,zhang_enhanced_2007,komarala_surface_2008,l._-viger_plasmon-enhanced_2011,ghenuche_nanophotonic_2014,lu_plasmon_2014,reil_forster-type_2008,zhao_plasmon-controlled_2012,marocico_spectral_2014,cortes2017super}, suppressed \cite{leitner_reduced_1988,kim_switching_2012,reil_forster-type_2008,zhao_plasmon-controlled_2012,marocico_spectral_2014,Tumkur_FRET2015}, or remain unchanged \cite{de_dood_forster_2005,blum_nanophotonic_2012,rabouw_photonic_2014,schleifenbaum_dynamic_2014}. While the theory of FRET has been developed extensively since the first initiation by F\"orster, ranging from semi-classical electrodynamic theories to microscopic and macroscopic quantum electrodynamic (QED) theories \cite{chance_comments_1975,agranovich1978electron,hua_theory_1985,kurizki_suppression_1988,john_quantum_1991,kobayashi_resonant_1995,bay_atom-atom_1997,agarwal_microcavity-induced_1998,klimov_resonance_1998,andrews1999resonance,basko_electronic_2000,dung_intermolecular_2002,des_francs_fluorescence_2003,govorov_theory_2007,durach_nanoplasmonic_2008,hohenester_interaction_2008,pustovit_resonance_2011}, there remains a significant disparity between experimental results and theoretical predictions. This has also led to conjectures that the FRET rate is independent of the photonic environment. Thus, no insightful approach exists that can explain the underlying physics behind the disparity of observations in experiments. 

\begin{table}[h]
\caption{Experimental Results of FRET}
\vspace{0.1cm}
\centering
\begin{tabular}{c|c|c|c}
 & Enhancement  & Suppression  & No effect  \\ \hline
Microcavity &\cite{hopmeier_enhanced_1999,andrew_forster_2000,finlayson_enhanced_2001,ghenuche_nanophotonic_2014} & \cite{Tumkur_FRET2015} & \cite{de_dood_forster_2005,blum_nanophotonic_2012,schleifenbaum_dynamic_2014} \\
Nanoparticles  &\cite{lakowicz_radiative_2002,fujiwara_enhancement_2005,zhang_enhanced_2007,komarala_surface_2008,l._-viger_plasmon-enhanced_2011,lu_plasmon_2014,reil_forster-type_2008,zhao_plasmon-controlled_2012,marocico_spectral_2014}&\cite{leitner_reduced_1988,kim_switching_2012,reil_forster-type_2008,zhao_plasmon-controlled_2012,marocico_spectral_2014} &\cite{rabouw_photonic_2014} \\
\end{tabular}
\end{table}

The purpose of the present work is to elucidate the fundamental nature of FRET in a nanophotonic environment for  experimentally-relevant scenarios, thereby resolving the debate that has been ongoing for fifteen years.
We show that the QED perturbative approach to analyzing dipole-dipole interactions in a nanophotonic environment can be completely captured by an effective near-field dipole model. This lends itself to a physically intuitive picture of environmental FRET rate modification that was not available before. As a result of our model, we are able to conclusively explain multiple recent experiments which have shown surprisingly contradictory results of enhancement, suppression and even no effect of the environment on FRET. We also show that the lack of FRET rate variation in recent experiments is not universal behaviour but is strongly dependent on environmental conditions and  orientational effects. We define a FRET rate enhancement factor analogous to the Purcell factor used in spontaneous emission modification calculations to quantify the influence of nanostructures on the FRET rate. We also introduce the concept of a two-point spectral density which quantifies the photonic modes mediating FRET in analogy to the photonic density of states which controls spontaneous emission. These introduced figures of merit (FOM)  captures the contradictory regimes of FRET completely in the widely used planar and spherical nanostructured geometries. Our work also clarifies an important puzzling observation that the FRET rate enhancement factor ($F_{ET}$) is in general much smaller than the Purcell factor ($F_{p}$) in most experiments; however, we show how careful design of environmental properties can result in cases where $F_{ET}$ can be greater than $F_p$. A striking manifestation of this property can be exploited to enhance FRET efficiency in an all-dielectric (transparent and lossless) system where we predict that the FRET efficiency can be enhanced by more than 300-400\%. Finally we place fundamental bounds on the achievable figures of merit in realistic photonic environments that arise from orientational effects and limitations of intrinsic quantum efficiency of donors.


\section{Theory of spontaneous emission rate vs FRET rate}
It is well understood that spontaneous emission and FRET have underlying connections but for completeness, we emphasize below that the environmental attributes which govern them are entirely different. 
Using Fermi's Golden rule and the first order transition amplitude, one arrives at the general expression of the spontaneous emission (SE) rate 
\begin{equation}
	\gamma_{D,rad} = \frac{2\omega_D^2|p_D|^2}{\hbar\epsilon_0 c^2}[\mathbf{n}_D \cdot \text{Im}\{\bar{\bar{G}}(\mathbf{r}_D,\mathbf{r}_D;\omega_D)\}\cdot \mathbf{n}_D ]
	\label{gamma o}
\end{equation}
for an atom in an inhomogeneous environment. $\omega_D$ is the radial frequency, $\hbar$ is the reduced Planck's constant, $c$ is the speed of light, $\epsilon_o$ is the free-space permittivity, and $\{\bar{\bar{G}}(\mathbf{r},\mathbf{r};\omega)$ is the classical dyadic Green function related to the electric field of the dipole. The atom is described by dipole moment $\mathbf{p}_D = p_D\mathbf{n}_D$ in the dipole approximation. It is well established that the SE rate is dependent on the density of photonic modes of the environment quantified by the partial local density of states (LDOS) \cite{novotny2012principles}
\begin{equation}
			\rho_E(\mathbf{r}_D;\omega) = \frac{6\omega}{\pi c^2} \mathbf{n}_D \cdot \text{Im}\{\bar{\bar{G}}(\mathbf{r}_D,\mathbf{r}_D;\omega)\}\cdot \mathbf{n}_D
	\label{LDOS}
\end{equation}
for a given dipole orientation specified by the unit vector, $\mathbf{n}_D$. Note that the LDOS depends on the position $\mathbf{r}_D$ of the donor dipole moment only. An enhancement in the LDOS $\rho_E(\mathbf{r}_D;\omega)$ compared to that in vacuum $\rho_E^o(\mathbf{r}_D;\omega)$ translates into a larger decay rate described by the Purcell factor
\begin{equation}
	F_p = \frac{\gamma_{D,rad}}{\gamma_{D,rad}^o}  = \frac{\rho_E(\mathbf{r}_D;\omega)}{\rho_E^o(\mathbf{r}_D;\omega)}.
\end{equation}
This is the well known figure of merit that is used to describe the decay rate enhancement or suppression of spontaneous emission in the case of low absorption. In the general case where the donor atom has an intrinsic quantum yield given by $Q_D = \frac{\gamma_{rad}}{\gamma_{rad} + \gamma_{nrad}}$, where $\gamma_{nrad}$ is a nonradiative decay rate that is assumed to be independent of the environment, then the overall enhancement of spontaneous emission is given by \cite{ford1984electromagnetic,chance1978molecular}
\begin{equation}
	\frac{\gamma_D}{\gamma_D^o} = (1-Q_D) + Q_D\,F_p.
\end{equation}

\begin{figure}[t!]
\begin{center}
\includegraphics[width=84mm]{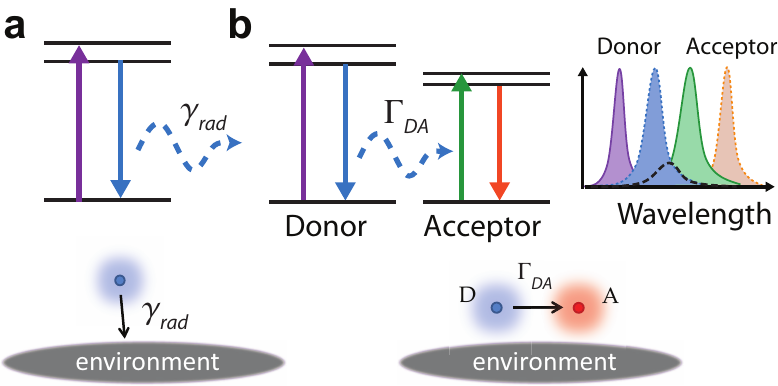}
\end{center}
\vspace{-10px}
\caption{(a) Energy-level diagram depicting spontaneous emission. $\gamma_{rad}$ denotes the rate of radiative energy transfer to any location in the environment. The acceptor is not considered as part of the environment. (b) Energy-level diagram depicting FRET. FRET occurs when two neighboring atoms or molecules, denoted as donor and acceptor, have overlapping emission and absorption spectra and couple due to a Coulombic dipole-dipole interaction. The FRET rate $\Gamma_{DA}$ denotes the energy transfer to the acceptor location only.
\label{figure1} }
\end{figure}

\paragraph{Environment-modified FRET.} While the preceding results have been well established in literature, questions have emerged whether the same physical quantities can characterize the effect of the environment on F\"orster resonance energy transfer. FRET can be understood as spontaneous emission of a donor molecule to the specific location of the acceptor triggered by near-field dipole-dipole interaction. Within the semi-classical picture, a donor dipole induces a dipole moment of a nearby polarizable acceptor. If the donor and acceptor have overlapping emission and absorption spectra (see Fig. \ref{figure1}(b)), then resonance energy transfer occurs. In the QED picture, FRET rate variation is mediated by virtual photons \cite{andrews1999resonance}. One arrives at the well-known expression for the FRET rate,
\begin{equation}
	\Gamma_{DA} = \frac{2\pi}{\hbar^2}\int\!d\omega \, |V_{EE}(\omega)|^2 \sigma_D(\omega)\sigma_A(\omega)
    \label{GammaDA}
\end{equation}
using Fermi's Golden rule by including the second order transition amplitude \cite{dung_intermolecular_2002}. Here, $\sigma_D(\omega)$ and $\sigma_A(\omega)$ represent the single-photon emission and absorption spectra of the donor and acceptor respectively. The results given above are applicable for quantum emitters placed in a linear, absorbing and dispersive electromagnetic environment. We emphasize that elaborate calculations with macroscopic QED theory yield the exact same results as a semi-classical theory with only minor differences in the definition of the dipole moment (see appendix). 

The dependence of the FRET rate $\Gamma_{DA}$ on the environment is clear through the resonant dipole-dipole interaction (RDDI)\cite{dung_intermolecular_2002},
\begin{equation}
	V_{EE}(\mathbf{r}_A,\mathbf{r}_D;\omega) = -\frac{\omega^2}{\epsilon_o c^2} \mathbf{p}_A\cdot\bar{\bar{G}}(\mathbf{r}_A,\mathbf{r}_D;\omega)\cdot\mathbf{p}_D
	\label{RDDI}
\end{equation}
as the environmental quantity that acts in an analogous way to the LDOS but instead mediates the magnitude of the FRET dipole-dipole interaction through virtual photon exchange. 
The subscript $_{EE}$ refers to an electric dipole interaction which is governed by the electric field of each dipole. The definition can be generalized to the case of magnetic dipole interactions ($V_{BB}$) and electric-magnetic dipole interactions ($V_{EB}$) as well. Note that in the near-field limit ($kr<<1$), the FRET rate reduces to the well known F\"orster formula ($\Gamma_{DA} \propto 1/n^4r^6$)  with the expected $r^{-6}$ and $n^{-4}$ dependence.

Note that the LDOS and FRET rate are both dependent on the dyadic Green function which contains all of the environmental information. The FRET rate is clearly not dependent on the LDOS, as debated in several papers \cite{blum_nanophotonic_2012,de_dood_forster_2005,rabouw_photonic_2014}. Nevertheless, it is dependent on the environment through the two-point Green function. While the LDOS is a measure of the energy transfer rate to any location in the environment, the RDDI is a measure of the energy transfer rate to the acceptor location only (see Fig.1). Analogous to the Purcell factor, we now introduce the FRET rate figure of merit
\begin{equation}
	F_{ET} = \frac{\Gamma_{DA}}{\Gamma_{DA}^o} 
\end{equation}
where the denominator is the homogeneous FRET rate. Combining these results, the total decay rate of a (high-yield) donor in an inhomogeneous environment is
\begin{equation}
	\gamma_{DA} = F_p \gamma_D^o + F_{ET}\Gamma_{DA}^o
\end{equation}
where the first term denotes the modified spontaneous emission rate and the second term denotes the modified FRET rate. The equation above is important for the rest of the discussion in the paper and is always valid in the weak-coupling regime of FRET (irreversible, incoherent energy transfer).

\begin{figure*}[t!]
\begin{center}
\includegraphics[scale = 0.71]{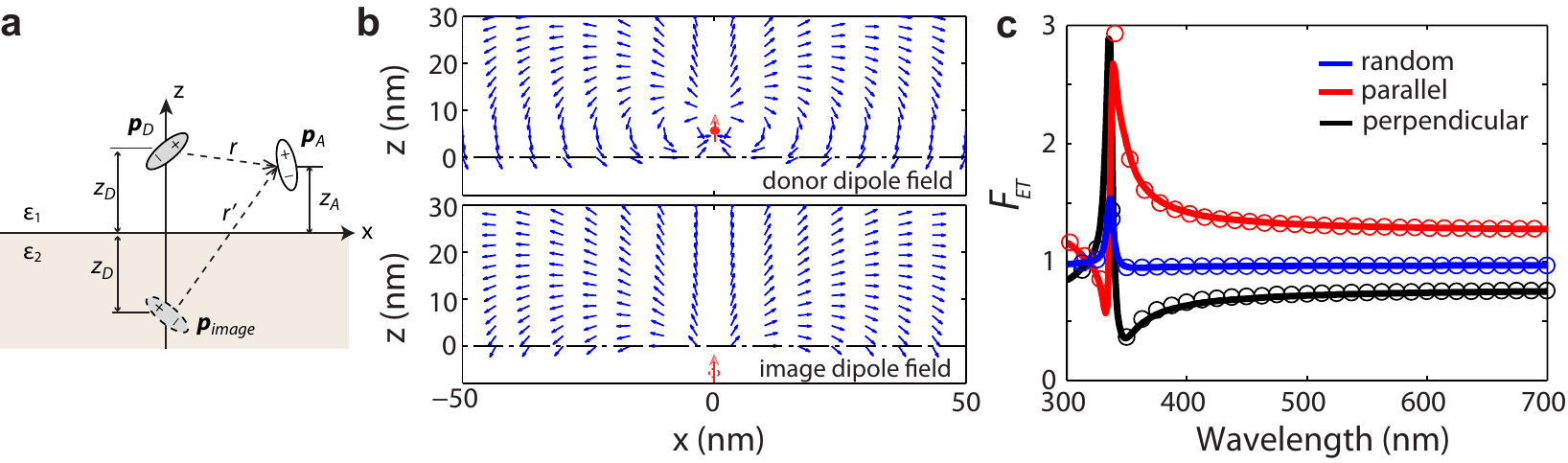}
\end{center}
\vspace{-10px}
\caption{\textbf{Environment modified dipole-dipole interactions and FRET.} QED theory of dipole-dipole interactions in the near-field is completely captured by an effective dipole model. FRET is governed by induced image dipoles in the metallic environment explaining a multitude of puzzling experimental observations. (a) Image dipole method for half-space structure. The magnitude of the image dipole moment is given by $p_{image} = \frac{\epsilon_2 - \epsilon_1}{\epsilon_2 + \epsilon_1}p_D$. (b) Visualization of normalized electric field plots for vertical donor dipole (above) and vertical image dipole (below) with $|\epsilon_2| > |\epsilon_1|$. Note that a non-trivial superposition of fields due to the vectorial nature of the electric field results in regimes of suppression, enhancement, and null effect on FRET. These regimes cannot be explained by the LDOS or Purcell factor alone. (c) FRET rate figure of merit for two dipoles 7 nm apart, and 7 nm above silver. Enhancement is seen when $|\epsilon_2| < |\epsilon_1|$, suppression is seen when $|\epsilon_2| > |\epsilon_1|$, while no effect is seen when $|\epsilon_2| \approx |\epsilon_1|$. These regimes are determined by the orientation of the image dipole. Note also that the FRET rate enhancement has a non-trivial dependence on the wavelength (see also Table 1). Exact QED results are denoted by the solid lines which are in complete agreement with our analytical expressions (circles).
\label{figure2} }
\end{figure*}

\section{Near-field image dipole model} 

We now show that the QED picture, consisting of virtual photon exchange, is completely captured by an effective dipole model that may be derived independently using the method of images. As shown in Fig. \ref{figure2}(a), we consider a half-space configuration where the donor and acceptor dipoles are embedded in the upper half-space medium described by the permittivity $\epsilon_1$. The lower half-space is described by permittivity, $\epsilon_2 = \epsilon_2' + i\epsilon_2''$, where $\epsilon_2'$ denotes the real part of the permittivity and $\epsilon_2''$ denotes the imaginary part of the permittivity. In a homogeneous environment, the FRET rate becomes comparable in magnitude to the SE rate for donor-acceptor separation distances of $r\ll c/\omega_D$. This implies that the dipole-dipole interaction is dominated by a quasistatic field. Therefore to control FRET we must engineer the \emph{quasistatic} fields as opposed to the electrodynamic fields. 
In both semi-classical and quantum theories of energy transfer, the FRET rate Eq. (\ref{GammaDA}) is governed by the projected electric field induced by the donor, $\mathbf{E}(\mathbf{r}_A;\omega) = \omega^2\mu_o \bar{\bar{G}}(\mathbf{r}_A,\mathbf{r}_D;\omega)\cdot\mathbf{p}_D$, at the location of the acceptor. In the near-field of a planar interface (see Fig. \ref{figure2}(a)), the projected electric field can be written as
\begin{equation}
	\mathbf{n}_A\cdot\mathbf{E}(\mathbf{r}_A;\omega_D) = \frac{p_D}{4\pi n_1^2}\frac{\kappa}{r^3} + \frac{p_{image}}{4\pi n_1^2}\frac{\kappa'}{r'^3}
	\label{ImageDipole}
\end{equation}
which simply denotes the dipole-dipole interaction between the donor and acceptor (first term) and the image dipole and acceptor (second term). As shown in Fig. \ref{figure2}(a), $r = |\mathbf{r}|$ and $r' = |\mathbf{r'}|$ represent the donor-acceptor and image-acceptor separation distances respectively, $\kappa=3\cos\theta_{Ar}\cos\theta_{Dr}-\cos\theta_{A,D}$ is the commonly used orientation parameter between the donor and acceptor \cite{andrews1999resonance}, and $\kappa'=3\cos\theta_{A{r'}}\cos\theta_{image,{r'}}-\cos\theta_{A,image}$ is the orientation parameter between the acceptor and the image dipole. Here, the angles are formally defined through the relations, $\mathbf{n}_D\cdot\mathbf{\hat r} = \cos\theta_{Dr}, \mathbf{n}_A\cdot\mathbf{\hat r} = \cos\theta_{Ar},  \mathbf{n}_A\cdot\mathbf{n}_D = \cos\theta_{AD}$, as well as $\mathbf{n}_A\cdot\mathbf{\hat r'} = \cos\theta_{Ar'}, \mathbf{n}_{image}\cdot\mathbf{\hat r'} = \cos\theta_{image,r'}, \mathbf{n}_A\cdot\mathbf{n}_{image} = \cos\theta_{A,image}.$

Equations (\ref{GammaDA}) and (\ref{ImageDipole}) imply that the total FRET rate is dependent on the total field generated at the acceptor location by the donor as well as its image dipole. This leads to a subtle interplay of interference effects that completely govern the nature of FRET. Note that while the donor dipole moment $p_D$ is independent of the environment, the induced image dipole has an environment-dependent dipole moment given by
\begin{equation}
	p_{image} = \frac{\epsilon_2 - \epsilon_1}{\epsilon_2 + \epsilon_1}p_{D}.
	\label{image}
\end{equation}
The FRET rate figure of merit is finally given by
\begin{equation}
	F_{ET} =  \left|\frac{\kappa}{r^3} + \frac{\epsilon_2 - \epsilon_1}{\epsilon_2 + \epsilon_1}\frac{\kappa'}{r'^3}\right|^2.
	\label{FRETfom}
\end{equation}
At this point we must make several observations in comparison to experimental results. First, note that the magnitude of the image dipole moment is enhanced at the surface plasmon (SP) resonance condition ($\epsilon_1 + \epsilon_2 = 0$) \cite{raether1988surface}. As expected, this explains why many experiments have observed FRET rate enhancements near the SP resonance (see Table 1). More importantly, FRET can be enhanced or suppressed depending on interference effects arising from the field of the image dipole. This is clearly evident from our theoretical model since the orientation of the image dipole determines constructive or destructive interference effects depending on the magnitude of $\epsilon_1$ and $\epsilon_2$. The image dipole maintains its original orientation [see Fig. \ref{figure2}(a)] if $|\epsilon_2| > |\epsilon_1|$ leading to enhancement, on the other hand, the image dipole orientation flips if $|\epsilon_2| < |\epsilon_1|$ causing FRET rate suppression (see Eqs. (\ref{ImageDipole}) and (\ref{FRETfom})). Thus in the case of metals which are highly dispersive, both regimes can be observed in different wavelength ranges.

Third, note that the vectorial structure of the near-field of the donor and image dipole results in a spatial inhomogeneity of the quasi-static fields. This affects the FRET rate depending on the acceptor location (see Fig. \ref{figure2}(b)). This explains why various donor-acceptor geometric configurations with similar materials can result in suppression or enhancement of FRET. We emphasize that two important aspects which govern the regimes of enhancement, suppression or no effect throughout the various experimental studies of FRET are (a) the spectral overlap region in which FRET takes place and (b) the donor-acceptor separation compared to the distance from the photonic environment. The FRET rate will be independent of the environment only if the distance of the donor-acceptor pair from the planar interface is much larger than the donor-acceptor separation distance. We also expect environmental effects to be significantly reduced when the mode resonances of the photonic environment lie outside the spectral overlap region of the donor-acceptor pair, as predicted by the perturbative formalism valid in the weak-coupling regime.

(1) \textit{Perfect Reflector}. For the case of a perfect reflection Eq. (\ref{image}) shows that the image dipole moment magnitude is equal to the donor dipole moment magnitude. The perfect reflector condition is satisfied when there is high absorption in medium 2 ($\epsilon_2''\rightarrow \infty$), or when there is a high contrast between the permittivities of medium 1 and 2 ($|\epsilon_2'|/\epsilon_1 \gg 1$) which typically occurs far from the SP resonance in the long-wavelength limit. 

A simple analysis of the FRET rate dependence yields the following table of regimes:
\begin{table}[h!]
\caption{Perfect reflector regimes for $F_{ET}$}
\centering
\begin{tabular}{c|c|c|c}
$F_{ET}$ & collinear dipoles   & coparallel dipoles   & randomly oriented   \\ 
      & ($\kappa^2=4$) & ($\kappa^2=1$) & ($\braket{\kappa^2}=2/3$)\\
\hline
$z_D/\rho \ll 1$ &\it suppression &\it enhancement &\it suppression \\
$z_D \sim \rho$  &\it enhancement &\it suppression &\it no effect  \\
$z_D/\rho \gg 1$ &\it no effect &\it no effect &\it no effect
\end{tabular}
\end{table}

In the table, we let $z_A=z_D$,  $\rho$ is the horizontal separation distance between donor and acceptor, and we have used $\kappa^2$ as the commonly used relative orientation parameter between donor and acceptor. Note the large sensitivity of the behaviour of FRET to both distance and orientation which has not been elucidated before. As outlined in Table 2,  the perfect metal reflector can surprisingly inhibit or enhance FRET and even have no effect depending on the orientations and distances achieved in experiment. 



(2) \textit{Realistic Metal}.
For the case of realistic losses, the FRET figure of merit across wavelengths has the form
\begin{equation}
	F_{ET} = \frac{[(\epsilon_2'+\epsilon_1)+q]^2}{|\epsilon_2 + \epsilon_1|^2} + \frac{\epsilon_2''^2(1+\Omega)^2}{|\epsilon_2 + \epsilon_1|^2}
	\label{FET quasi}
\end{equation}
where $q=\Omega(\epsilon_2'-\epsilon_1)$ is a Fano-like parameter and $\Omega=(r^3/r'^3)(\kappa'/\kappa)$ is an orientation/distance parameter. The first term, which corresponds to a dispersive dipole-dipole interaction, resembles a Fano resonance profile. The second term which depends on $\epsilon_2''$ is a dissipative dipole-dipole interaction with a Lorentzian-like profile. When losses are sufficiently low the Fano term dominates. The result is shown in Fig. 2(c), which shows the excellent agreement between the exact QED result (solid line in Fig. 2(c)) and the quasistatic expression [Eq. (\ref{FET quasi})]. The same system can show regions of enhancement, suppression, and no effect depending on the wavelength of operation. These various regimes are explained by the image dipole model, as outlined in the perfect reflector case. 

In stark contrast, the Purcell factor for a realistic metal is approximated by
\begin{equation}
	F_{p} = 1 + \frac{3}{16}\frac{\epsilon_1\epsilon_2''}{|\epsilon_1+\epsilon_2|^2}\frac{\kappa'}{(k_1 z_D)^3}
	\label{Fp quasi}
\end{equation}
which is always enhanced in the near-field ($F_p>1$).

We now contrast the fundamental differences between the Purcell factor and FRET rate enhancement factor at the SP resonance ($\epsilon_2'=-\epsilon_1$). Equation (\ref{FET quasi}) reduces to
\begin{equation}
	F_{ET} = 4\frac{\kappa'^2r^6}{\kappa^2r'^6}\left(\frac{\epsilon_1}{\epsilon_2''}\right)^2 + \left(1+\frac{\kappa'r^3}{\kappa r'^3}\right)^2.
	\label{FET reson}
\end{equation}
where we identify the quantity $Q=\epsilon_1/\epsilon_2''$ as the quality factor of the SP resonance.  Similarly, Eq. (\ref{Fp quasi}) reduces to
\begin{equation}
	F_p = 1 + \frac{3}{32\pi^3n_1^3}\left(\frac{\lambda_D}{z_D}\right)^3\left(\frac{\epsilon_1}{\epsilon_2''}\right)
	\label{Fp reson}
\end{equation}
at the SP resonance. Note these expressions separate the material and geometrical properties of the system. Furthermore, the Purcell factor follows the well-known linear dependence on the quality factor while $F_{ET}$ is dependent on the square of the quality factor. The proportionality of the quality factors follows from the linear dependence and squared magnitude dependence on the Green function for the spontaneous emission and energy transfer rates respectively.

\begin{figure}[t!]
\begin{center}
\includegraphics[width=70mm]{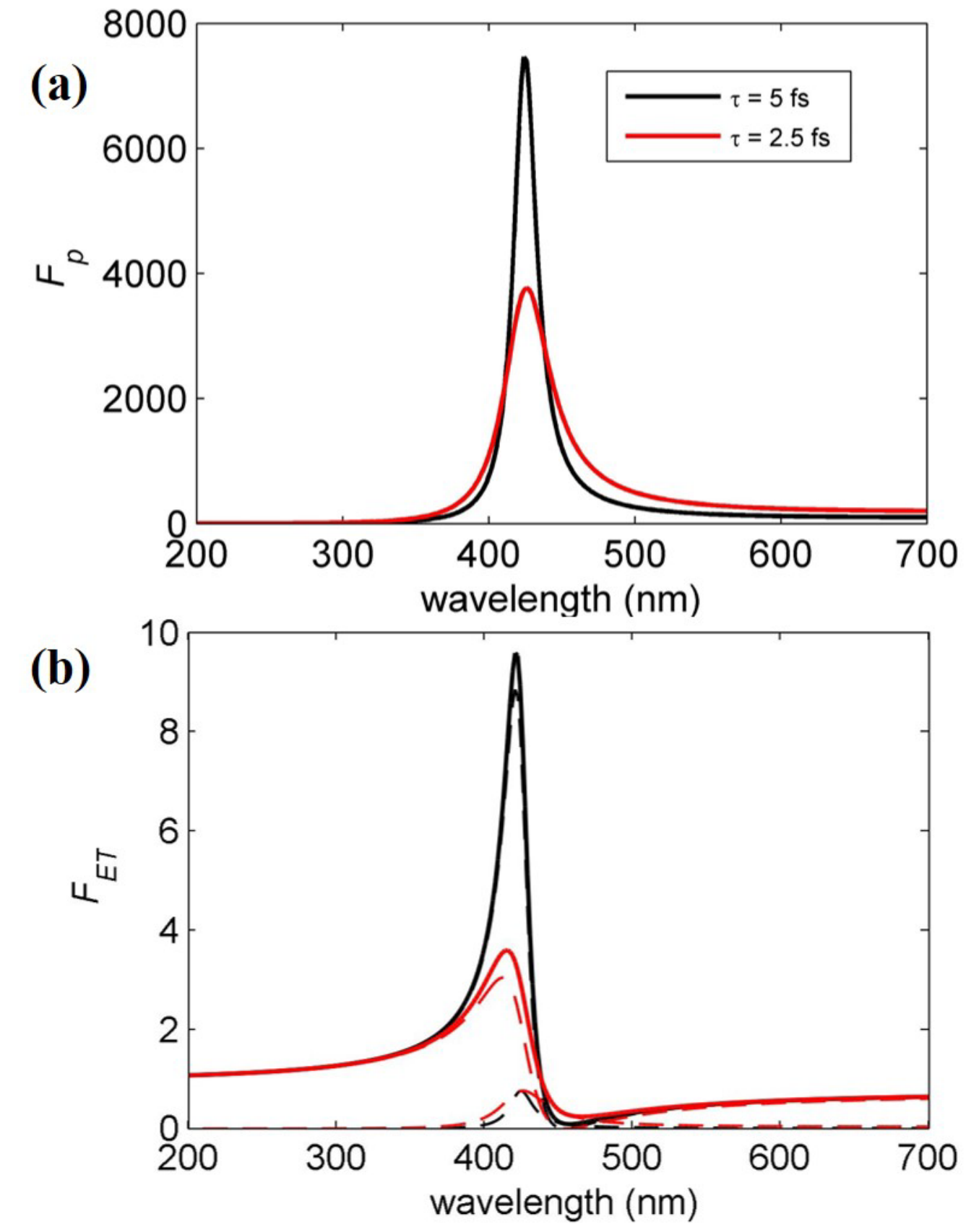}
\end{center}
\vspace{-10px}
\caption{\textbf{Effect of losses}. \textbf{(a)} Purcell factor $F_p$. \textbf{(b)} FRET figure of merit $F_{ET}$. Bottom half-space is modelled as Drude metal with $\omega_p = 6.3\times10^{15} s^{-1}$ and the Drude relaxation time of $\tau=5 fs$ (black) and $\tau=2.5 fs$ (red). Dashed lines correspond to the two terms, dispersive dipole-dipole interaction and dissipative dipole-dipole interaction, in Eq. (\ref{FET quasi}). Note the FRET enhancement factor is in general much smaller than the Purcell factor in agreement with widely reported observations.
}
\end{figure}

We now use Eqs. (\ref{FET reson}) and (\ref{Fp reson}) to shed light on why Purcell factor enhancements have always shown large values in comparison to FRET rate enhancements (see Table 1). 
The key distinction between the two figures of merit lies in the length scales governing them captured by the $r^6/r'^6$ and $\lambda_D^3/z_D^3$ terms. For the half-space problem that we have considered here, geometric considerations imply that the donor-acceptor distance ($r$) is always smaller than the image-acceptor separation ($r'$). This drastically reduces the FRET figure of merit due to the sixth power dependence. The length scale factor of spontaneous emission, on the other hand, depends on the ratio of the wavelength of emission to the emitter distance from environment ($\lambda_D^3/z_D^3$) and is always much larger than 1 in the near-field. The cubed power dependence helps enhance the overall Purcell factor and so in general we find that regardless of the material quality factor, the FRET figure of merit will be much smaller than the Purcell factor for realistic metals. In Fig. 3, we compare the Purcell factor and FRET enhancement factors across wavelengths which re-emphasizes this important point. Note that this explains the multitude of experiments which have observed negligible enhancements of the FRET rate as opposed to the SE rate (See Table 1).

(3) \textit{Dielectric}. We now consider medium 2 as a dielectric ($\epsilon_2>0$). In this case, there are three general scenarios that may occur. (i) If $\epsilon_1 \approx \epsilon_2$ then the magnitude of the image dipole moment approaches zero and $F_{ET}\approx 1$. (ii) If $\epsilon_2/\epsilon_1 \gg 1$ then the dielectric acts like a perfect reflector with the different regimes outlined in the previous section. (iii) If $\epsilon_2/\epsilon_1 \ll 1$ then the image dipole flips orientation and the dielectric acts like a perfect reflector with opposite regimes to those outlined in Table 2.
\newline

\textbf{FRET efficiency}. We now place limits on the FRET efficiency based on the quantum yield of the donor and elucidate fundamental competition between FRET efficiency and the Purcell factor. 
The efficiency of energy transfer to the acceptor location as compared to the energy transfer to the rest of the environment is given by
\begin{equation}
	\eta = \frac{\Gamma_{DA}}{\gamma_D + \Gamma_{DA}}.
	\label{FRET efficiency}
\end{equation}
In many applications and experiments, controlling the FRET efficiency is as important as controlling the FRET rate hence we introduce the FRET efficiency figure of merit:
\begin{equation} \label{Feff}
F_{eff}  = \frac{\eta}{\eta_o} 
  = \frac{F_{ET}}{F_{ET}\eta_o + [(1-Q_D) + Q_DF_p](1-\eta_o)}
\end{equation}
where $\eta_o$ denotes the FRET efficiency in a homogeneous environment. It then follows that the condition of FRET efficiency enhancement ($F_{eff}>1$) is $F_{ET} > (1-Q_D) + Q_D\, F_p$ which shows that the intrinsic quantum efficiency of donors ($Q_D$) has a large effect on the FRET efficiency. 
\newline

\textit{(1) High-yield donor ($Q_D \approx 1$)}. From the above equations, the condition to increase the FRET efficiency for high yield donors requires the FRET rate enhancement factor to be larger than the Purcell factor.
\begin{equation}
F_{eff} > 1 \implies F_{ET}>F_{p}
\end{equation}
Our analysis from the previous section shows that this condition is very difficult to achieve for realistic metals. Using Eqs. (\ref{FET reson}) and (\ref{Fp reson}), we find that the minimum quality factor required to observe ($F_{ET} > F_p$) is given by
\begin{equation}
	Q > \frac{3}{128\pi^3n_1^3}\left(\frac{\lambda_D}{z_D}\right)^3\frac{\kappa^2r'^6}{\kappa'^2r^6}..
\end{equation}
For a  donor-acceptor pair equidistant from the metal surface, the minimum quality factor can range from $10^3$ and upwards. Since this value is not found in realistic metals, then it is generally the case that FRET efficiency cannot be enhanced with high-yield donors explaining observations of experiment.


\textit{(2) Low-yield donor}. Using realistic parameters for metal-based systems (e.g $F_p\approx 10$ and $F_{ET}\approx 4$), we find that the intrinsic quantum efficiency must be less than 33\% in order to observe FRET efficiency enhancement. This implies that low quantum yield donors will exhibit an increase in the FRET efficiency even if $F_p > F_{ET}$.  We emphasize that this explains why many plasmonic-based experiments with metallic nanoparticles have observed enhancements in the FRET efficiency \cite{zhang_enhanced_2007,lunz_surface_2011}. However, note that if $F_p \gg F_{ET}$ then there will no efficiency enhancement even with a low quantum yield donor.

\textbf{FRET near nanosphere.} We now turn our attention to the effect of FRET near spherical nanoparticles, as has been the focus of a wide range of experiments [see Table 1]. Our effective near-field dipole model captures the observed effects of FRET near these structures, in full agreement with the QED result.  In the quasistatic regime, $kR\ll 1$, the nanoparticle can be treated as a dipole-driven multipolar source that acts to modify the overall FRET rate. For accurate predictions, we have used the full dyadic Green function for spherically multilayered media originally developed in \cite{li1994electromagnetic} (see Methods).

\begin{figure}[t!]
\begin{minipage}[b]{1\linewidth}
\begin{center}
\includegraphics[width=100mm]{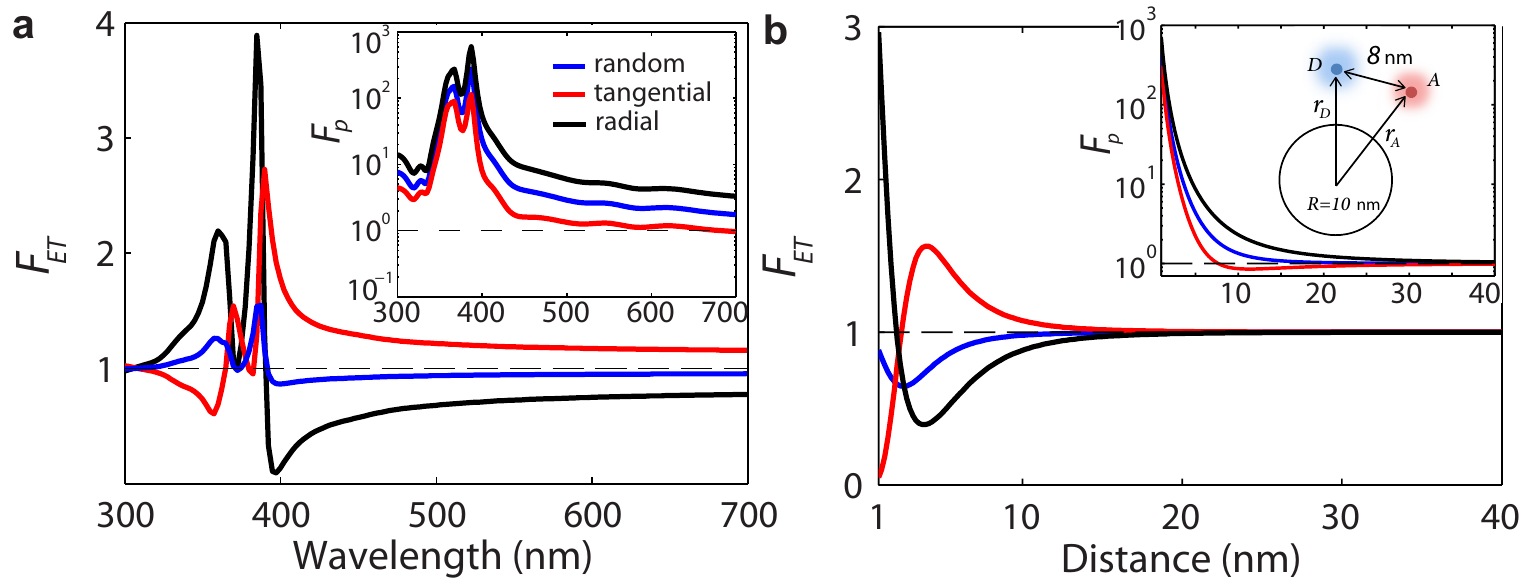}
\end{center}
\end{minipage}
\vspace{-5px}
\caption{\textbf{FRET near nanosphere.}  \textbf{(a)} FRET rate enhancement factor for spherical nanoparticle systems widely used in experiment. The donor and acceptor are both 8 nm away from an $Ag$ nanosphere of $10$ nm radius. Inset: Calculated Purcell factor for same system. The peaks are related to dipolar surface plasmon resonance and higher order multipolar non-radiative modes. We emphasize that $F_p \gg F_{ET}$ for plasmonic systems near the LSP resonance implying the energy transfer to the sphere (environment) is larger than the energy transfer to the acceptor. \textbf{(b)} Distance dependence of $F_{ET}$ and $F_p$ at the 650 nm wavelength region (away from resonance). Note that a tangential dipole exhibits a suppression in the Purcell factor due to near-field interference effects. This effect can be used to boost the FRET efficiency ($F_{eff} \propto F_{ET}/F_p$). The enhancement, suppression and null effect features in the three curves of different colors corresponding to the orientations of the dipole moments of the acceptor and donor are in agreement with Table 1.
\label{figure4} }
\end{figure}


In fig. \ref{figure4}(a), we compare the FRET rate enhancement factor with the Purcell factor for a donor-acceptor pair that has a fixed separation distance of 8 nm across the visible wavelength region. We consider a silver nanoparticle with a 10 nm radius. Two distinct peaks are observed in the spectrum. The peak at lower frequencies is a result of the dipolar surface plasmon resonance of the nanoparticle, while the second peak at higher frequencies is a result of the higher order (non-radiative) modes in the nanoparticle. Note that the Purcell factor ($F_p$) is orders of magnitude larger than the FRET rate enhancement factor ($F_{ET}$) near these resonant regions, in agreement with our previous arguments. Note, however, that in the low-frequency region the two factors became comparable in magnitude. Most interestingly, when the donor and acceptor are tangential to the surface of the sphere, the FRET rate is enhanced ($F_{ET}>1$) while the Purcell factor is suppressed ($F_{p}<1$). 


In fig. \ref{figure4}(b), we compare $F_{ET}$ and $F_p$ as a function of separation distance from the nanoparticle for an operating wavelength of 650 nm. Note that unlike the planar half-space case, the Purcell factor of a tangential dipole can be suppressed for certain distances. We emphasize that the FRET characteristics from the effective dipole model given in Table II  captures the distance and orientational dependence completely even in this spherical geometry.

\begin{figure}[t!]
\begin{minipage}[b]{1\linewidth}
\begin{center}
\includegraphics[width=100mm]{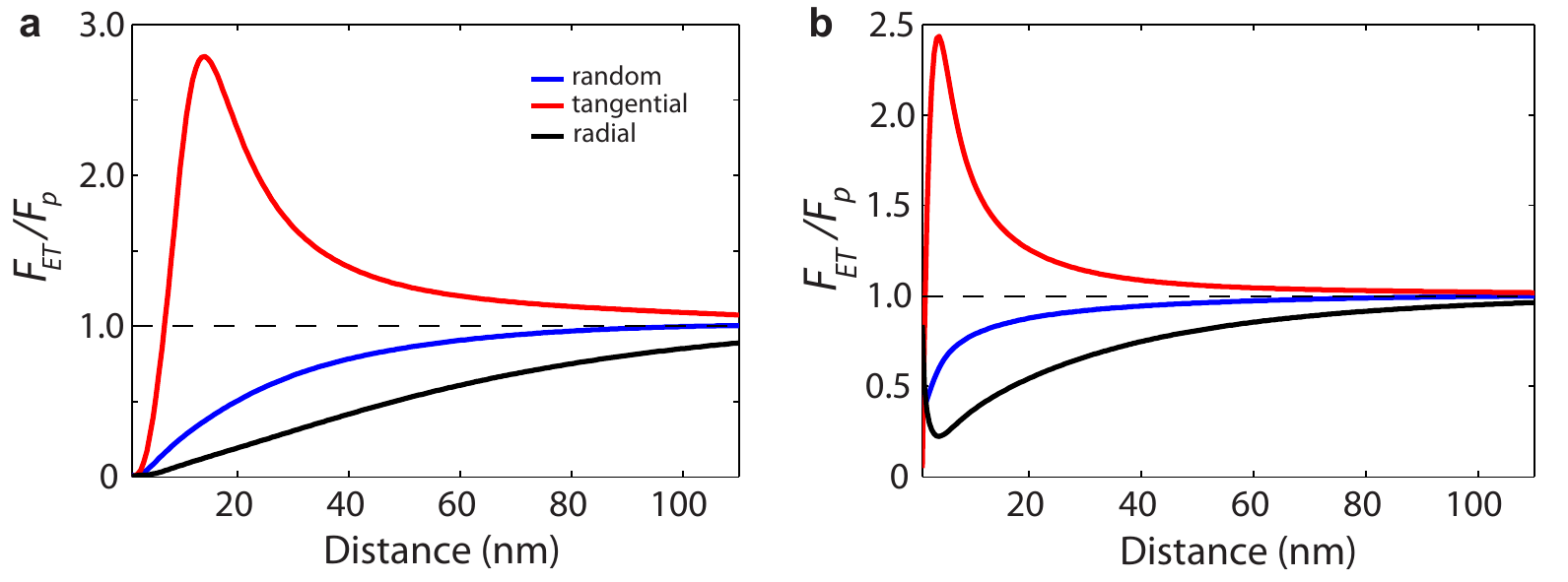}
\end{center}
\end{minipage}
\hspace{-4px}

\caption{\textbf{FRET efficiency.} \textbf{(a)} FRET efficiency enhancement occurs when $F_{ET}/F_p > 1$. We show that this ratio can be optimized for particular distances away from the nanoparticle. Results are shown for same set-up as Fig. \ref{figure4}(b) but with $R=40$ nm nanoparticle, where $\epsilon_1 = (1.33)^2$ and $\epsilon_2 = -19.5 + 0.47i$. \textbf{(b)} Counter-intuitive to prevalent designs, here we provide an all-dielectric  design to engineer FRET efficiency using a transparent nanosphere ($\epsilon_2 = 6.25 > 0$) and 40 nm radius. The efficiency enhancement  in FRET implies a larger fraction of the donor energy is transferred to the acceptor in presence of the nanosphere. This effect arises from suppression of the Purcell factor which is necessary to avoid energy transfer to the environment.
\label{figure5} }
\end{figure}

In fig. \ref{figure5}, we directly compare the ratio of $F_{ET}$ and $F_p$ which dictates the FRET efficiency enhancement as outlined in the previous section. We consider the case shown in the inset of Fig. \ref{figure4}(b), but with a nanoparticle with 40nm radius. Note that in the quasistatic regime, a larger  radius enhances all of the figures of merit since they are directly related to the polarizability of the nanoparticle. This can be seen from the dipole contribution of the nanoparticle which has a polarizability of $\alpha = 4\pi\epsilon_o R^3 \frac{(\epsilon_2 - \epsilon_1)}{(\epsilon_2 + 2\epsilon_1)}$. In Fig. \ref{figure5}(a), we find that there exists an optimum separation distance where $F_{ET} \approx 1.06, F_p \approx 0.38$ yielding the ratio $F_{ET}/F_p \approx 2.79$ which occurs only for the case where the donor and acceptor have dipole moments oriented tangential to the spherical nanoparticle (co-tangential case, solid red line). We therefore suggest the implementation of co-tangential dipoles as an important design principle for enhancing the FRET efficiency in future experiments. 

In fig. \ref{figure5}(b) we present an all-dielectric platform for enhancing the FRET efficiency. We consider a dielectric nanosphere with $\epsilon_2 = 6.25$ and 40 nm radius. Here, the ratio is shown to be as large as $F_{ET}/F_p \approx 2.45$. We note that the overall enhancement of the efficiency is mainly due to the suppression of the Purcell factor $F_p\approx 0.48$ and a moderate enhancement in the FRET rate $F_{ET}\approx 1.2$. We emphasize that while the FRET rate can be drastically enhanced near resonances with large quality factors, the Purcell factor will be simultaneously enhanced as well. As a result, enhancing the FRET efficiency will be difficult to achieve near resonances since most of the energy from the donor is funneled to the environment (nanoparticle) and not the acceptor. However as we have shown in Fig. \ref{figure5}, away from resonances we find that engineering the FRET efficiency can result from the modification of the quasistatic fields. In essence, the nanosphere platform allows a simultaneous suppression of the Purcell factor while also enhancing the FRET rate close to the nanoparticle. Moreover, the material parameters largely become irrelevant away from resonance and in fact the orientational and the geometrical parameters play a much more important role.

\begin{figure*}[t!]
\begin{minipage}{1\linewidth}
\begin{center}
\centering
\includegraphics[height = 40mm, width = 120mm]{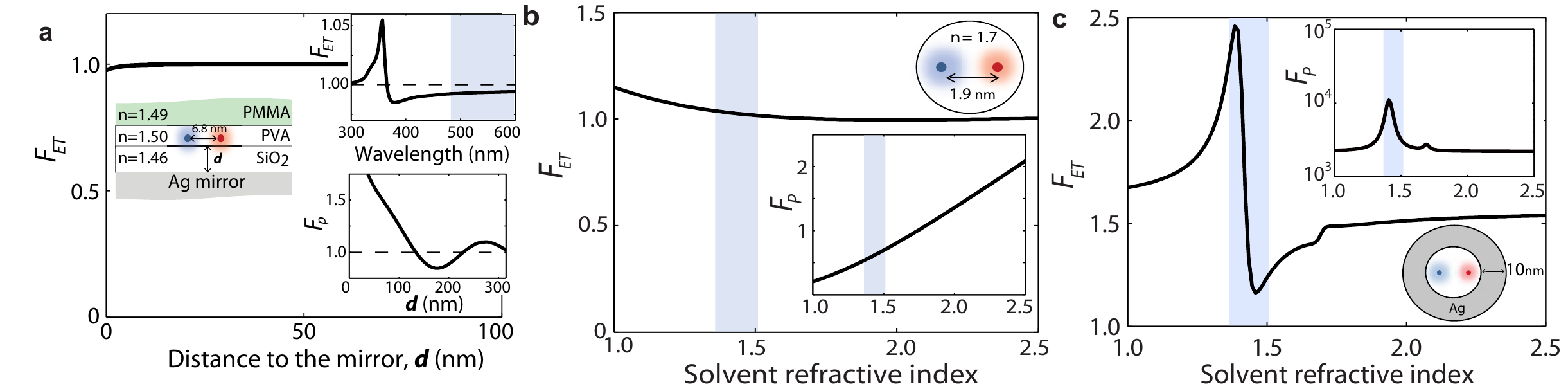}
\end{center}
\end{minipage}
\caption{\textbf{Comparison to experiments.} \textbf{(a)} Theoretical comparison to experiment in ref.\cite{blum_nanophotonic_2012}. The system configuration is shown in the inset. The FRET figure of merit is theoretically calculated to be $F_{ET}\approx 1$ for a wide range of separation distances $\bf d$ from the mirror, in agreement with the experiment (plotted at the donor's peak emission wavelength of 525 nm). Theoretical Purcell factor $\it F_p$ shows excellent agreement with experimental results (lower inset). However, using our theoretical model, we predict a drastic change in the FRET FOM near the $Ag$ SPP resonance in the limit $\textbf{d} \rightarrow 0$ (top inset). This shows that FRET rate can be modified for the same experiment if the regime is modified. \textbf{(b)} Theoretical comparison to experiment in \cite{rabouw_photonic_2014}. The donor-acceptor pair is embedded inside a nanocrystal (4 nm diameter) with assumed refractive index $n=1.7$ ($LaPO_4$). By varying the refractive index of the surrounding medium, we find that $F_{ET}\approx 1$ in agreement with our analysis. Note that we also predict the linear dependence of the Purcell factor as measured in the experiment (inset). \textbf{(c)} However, we predict that a silver-coated nanocrystal would produce a drastic change in the FRET FOM as well as the Purcell factor. This result would require the donor-acceptor overlap spectrum to lie around the 400 nm wavelength range. Note that the above results clearly show that FRET can be engineered by the environment even though it is extremely difficult in comparison to modifying spontaneous emission. The dyadic Green function formalism and results from QED theory were used to calculate all results and parameters were obtained from the experiments.
\label{figure6} }
\end{figure*}

\section{Comparison to experiments}

To provide a settlement to the debate surrounding FRET, we use our theory to explain recent experiments which have found that FRET intriguingly is not affected by the environment. Our goal is to show that this is not universal behavior. Our theory explains the experimental results while simultaneously pointing to regimes in the same experiment where FRET rate enhancement or suppression can occur. We have carefully isolated the experimental parameters of interest from the relevant works for theoretical consideration below.

We first turn to the planar multilayer system considered in \cite{blum_nanophotonic_2012} which showed a null result of the environment on FRET. Our theory shows that this result is specific to the experimental parameters considered. The multilayer structure is shown in the inset of Fig. \ref{figure6} consisting of a half cavity system. The donor-acceptor pairs are separated by a distance $d$ from a $Ag$ mirror. In the experiment, the distance from the mirror is varied in order to study the environmental influence on FRET. By using a transfer matrix method as well as the Green function formalism, we calculated the FRET figure of merit for the exact configuration in the experiment. Here, we show that our results capture the Purcell factor variation and match exactly with the observed null effect in the FRET rate. There are two reasons contributing to the no-effect result: (i) the separation distance between the donor-acceptor pair and the mirror is far too large to influence the quasistatic fields, and (ii) the surface plasmon resonance lies near the UV region which is outside the spectral overlap region of the donor-acceptor pair.  We show that if the same experiment is repeated with donor-acceptor spectra in the UV region, the SPP resonance can surely have an effect on FRET. This plasmonic resonance could be tailored with surface layers or metamaterials to overlap with the relevant spectrum for FRET.

We also conclusively explain a more recent experiment \cite{rabouw_photonic_2014} consisting of a donor-acceptor pair inside a $LaPO_4$ nanocrystal with $4$nm diameter (see inset Fig. \ref{figure6}(b)) which again showed the null result of the environment on the FRET rate. A change in the environment was achieved by dispersing the nanocrystals in solvents with different refractive indices. The range of the refractive index change in the experiment is highlighted in the blue region. Our theory predicts that for the exact range of solvent refractive indices considered in the experiment, FRET is unchanged in agreement with experiment. We note that our calculations for the SE rate enhancement also shows the same linear dependence that was observed in the experiment (see Fig. \ref{figure6}(b) inset) reinforcing our results. However, closer inspection reveals that changing the refractive indices beyond this range should have a clear effect on FRET. In fig. \ref{figure6}(c), we propose a new experimental set-up where the nanocrystal has a 10 nm silver coating. We have calculated the results for a 400nm operating wavelength to elucidate the role of resonances. It is striking that a subtle change in the exact same environment leads to a large variation in the FRET enhancement factor across the identical solvent index range that was probed in the experiment.

We now consider the FRET-nanoparticle experiment of \cite{zhang_enhanced_2007} which consisted of a donor-acceptor pair with a fixed separation distance of 9.2 nm placed at a fixed position away from a silver nanoparticle with 10 nm radius. Our results help to identify the orientational dependence of the observations and also point to mechanisms beyond conventional weak-coupling QED FRET theory. Based on the single-exponential lifetime fits from the experiment, the authors measured a donor lifetime enhancement of $\gamma_D/\gamma_D^o = 2.35$, a FRET rate enhancement of $\Gamma_{DA}/\Gamma_{DA}^o = 29.4$ and an overall FRET efficiency enhancement of $F_{eff} = 5.3$. The donor molecule had a peak emission wavelength at 662 nm. The trends of the theoretical figures of merit will be very similar to those in Fig. \ref{figure4}(b). For our simulation, we assumed that the donor and acceptor were both 4.5 nm away from the nanoparticle. We provide the overall results in the following table for various orientations.
\begin{table}[h]
\caption{Comparison to Lakowicz experiment for various orientations: FRET rate enhancement factor, Purcell factor, intrinsic quantum yield, FRET efficiency.}
\centering
\begin{tabular}{l|c|c|c|c}
           & $F_{ET}$ & $F_p$ & $Q_{D}$ & $F_{eff}$\\ \hline
co-tangential & 1.58     & 3.8  &  0.48   & 1.13\\
co-radial     & 0.39     & 13.9 &  0.11   & 0.15\\
random     & 0.72     & 7.2  &  0.22   & 0.41\\
\end{tabular}
\end{table}
In the table, $F_{ET}$ and $F_p$ were calculated independently based on the parameters mentioned above.  The intrinsic quantum yield $Q_D$ is used as a fitting parameter to reproduce the experimental decay rate enhancement $\gamma_D/\gamma_D^o = 2.35$. The theoretical FRET efficiency enhancement factor can be obtained using Eq. (\ref{Feff}) with an initial FRET efficiency of $\eta_o=0.12$,  given in \cite{zhang_enhanced_2007}. As we can see from the table, the co-tangential FRET case is the only one to reproduce the FRET efficiency enhancement seen in the experiment. Since the radial and random cases are in extreme disagreement with the experimental results, this might suggest that the measured lifetime values consisted primarily of the co-tangential dipole contribution. Such preferred orientations can arise from steric effects of ligands that attach the donor and acceptor to the nanoparticle. We must emphasize that the low quantum yield of the donor is an important factor in the observed FRET efficiency enhancement -- in agreement with our previous arguments. Moreover, while the co-tangential case provides qualitative agreement with the experiment, the theoretical values of $F_{ET}=1.58$ and $F_{eff}=1.13$ vastly underestimate the experimental values of $\Gamma_{DA}/\Gamma_{DA}^o = 29.4$ and $F_{eff}=5.3$ reported in the experiment. Since our current theory provides the upper bound for a weak-coupling model, these experimental results are therefore suggestive of more exotic physical phenomena that might be occurring. The possibilities include surface roughness induced plasmonic hot-spots that increase the two-point spectral density or coherence in energy transfer \cite{lee2007coherence}.

At this point, it is then important to consider some of the approximations that were made in the QED derivation of the FRET rate. The two primary approximations consists of: (1) weak-coupling and Markovian approximation, and (2) the dipole approximation. Since going beyond these approximations is beyond the scope of the current work, it would be interesting to see in the future how the transition to the strong coupling regime affects FRET in inhomogeneous environments, as well as how finite-size emitters affects FRET. The fundamental bounds we have established of FRET efficiency in the weak-coupling regime are extremely important to identify new pathways of energy transfer such as quantum coherence in photosynthesis \cite{lee2007coherence}.

\section{Conclusion}

 Our work contrasts the two-point spectral density which characterizes FRET analogous to the local density of states which characterizes spontaneous emission. We have clearly isolated the regimes which show enhancement, suppression and no effect on the FRET rate settling the long-standing debate surrounding FRET. Finally, we also conclusively explained several recent experiments that examined the role of the environment of FRET.  Engineering FRET is fundamental to multitude of applications from energy harvesting to molecular sensing and our theory provides intuitive insight unavailable up until now.

\section*{Appendix}
Here we show how the semi-classical and QED approaches result in the same expression for the environmental dependence of FRET. 

\textbf{Semi-classical approach.} From Poynting's theorem, the time-averaged power transfer from the donor to acceptor dipole is given by
\begin{equation}
	P_{DA} = \frac{\omega_o}{2}\text{Im}\{\mathbf{p}_A^*\cdot \mathbf{E_D(\mathbf{r}_A})\}.
	\label{PDA}
\end{equation}
The polarizable acceptor has induced dipole moment $\mathbf{p}_A = \bar{\alpha}_A \mathbf{E_D(\mathbf{r}_A})$. The electric field from a donor dipole evaluated at the location of the acceptor dipole $\mathbf{r}_A$ is 
\begin{equation} \label{ClassicEfield}
	\mathbf{E_D(\mathbf{r}_A}) = \frac{\omega_o^2}{\epsilon_0 c^2} \bar{G}(\mathbf{r}_A,\mathbf{r}_D;\omega_o)\cdot \mathbf{p}_D
\end{equation}
where $\bar{G}$ is the classical dyadic Green function that satisfies the inhomogeneous Helmholtz equation $
	 \left[ \frac{\omega^2}{c^2} \epsilon(\mathbf{r},\omega) - \nabla\times\nabla\times \right] \bar{G}(\mathbf{r},\mathbf{r}';\omega) = -\delta(\mathbf{r}-\mathbf{r}') $
together with the boundary condition at infinity. Combining Eqs. (\ref{ClassicEfield}) and (\ref{PDA}) gives the final result
\begin{equation}\label{PDAPo}
	\frac{P_{DA}}{P_o} = 6\pi\frac{\omega_o}{\epsilon_0c}\alpha_A''(\omega_o)|\mathbf{n}_A \cdot \bar{G}(\mathbf{r}_A,\mathbf{r}_D;\omega_o)\cdot \mathbf{n}_D|^2.
\end{equation}
Note that (\ref{PDAPo}) has been normalized by the total radiated power of the donor dipole $P_o =\frac{|\mathbf{p}_D|^2 \omega_o^4}{12\pi\epsilon_0 c^3}$. This is the main semi-classical result assuming that the donor dipole has a delta function emission spectrum.

\textbf{QED approach}. We consider the resonant energy transfer between donor $D$ and acceptor $A$ at position $\mathbf{r}_D$ and $\mathbf{r}_A$. The initial and final states are specified as
\begin{eqnarray}
	\ket{i} &= \ket{D^*,A}\otimes\ket{\{0\}} \nonumber \\ 
	\ket{f} &= \ket{D,A^*}\otimes\ket{\{0\}} 
\end{eqnarray}
where $^*$ denotes the excited state of the respective atom, and $\ket{\{0\}}$ is the vacuum state of the electromagnetic field. The transition rate between the initial and final state is given by  Fermi's Golden rule
\begin{equation}
	\Gamma_{i\rightarrow f} = \frac{2\pi}{\hbar}|M_{fi}|^2\delta(E_f-E_i).
\end{equation}
The matrix element $M_{fi}$ for a process undergoing a transition from $\ket{i}\rightarrow\ket{f}$ is given by
\begin{equation} \label{Matrix}
	M_{fi} = \braket{f|H_{int}|i} + \sum_k \frac{\braket{f|H_{int}|k}\braket{k|H_{int}|i}}{E_i - E_k}
\end{equation}
expanded up to second order. The summation over $k$ denotes the summation over all possible intermediate states. The interaction Hamiltonian, using the dipole approximation in the multipolar-coupling scheme, is  
\begin{equation} \label{Hint}
	\hat{H}_{int} = - \mathbf{\hat p}_D\cdot \mathbf{\hat E}(\mathbf{r}_D) - \mathbf{\hat p}_A\cdot \mathbf{\hat E}(\mathbf{r}_A)
\end{equation}
where $\mathbf{\hat p}_{D/A}$ is the donor/acceptor electric dipole moment operator and $\mathbf{\hat E}(\mathbf{r}_{D/A})$ is the quantized electric field evaluated at the donor/acceptor position.

The quantization of the electromagnetic field in dispersive and absorbing media is based on the introduction of noise charges and currents in Maxwell's equations, which give rise to Langevin equations that satisfy the fluctuation-dissipation theorem. In Fourier space, the electric field is related to the dyadic Green function by
\begin{equation}
	\mathbf{\hat E}(\mathbf{r},\omega) = \frac{i\omega}{\epsilon_oc^2} \int \! d^3\mathbf{r}'\bar{G}(\mathbf{r},\mathbf{r}',\omega)  \mathbf{\hat j}_N(\mathbf{r}',\omega).
\end{equation}
where the noise current operator is
\begin{equation}
	\mathbf{\hat j}_N(\mathbf{r},\omega) = \omega\sqrt{\frac{\hbar\epsilon_o}{\pi} \epsilon''(\mathbf{r},\omega)}\,\mathbf{\hat f}(\mathbf{r},\omega)
\end{equation}
and $\mathbf{\hat f}(\mathbf{r},\omega)$ denotes the bosonic field annihilation operator of the matter-assisted EM field. The Fourier space electric field is then related to a time-independent electric field of Eq. (\ref{Hint}) by
\begin{equation}
	\mathbf{\hat E}(\mathbf{r}) = \int_0^\infty\!\!\!\!\! d\omega \, \mathbf{\hat E}(\mathbf{r},\omega) + h.c.
\end{equation}
where $h.c.$ denotes the hermitian conjugate. 

Upon susbstitution of $H_{int}$ in Eq. (\ref{Matrix}) and simplification (details are given in \cite{dung_intermolecular_2002}), we find that the energy transfer rate is given by
\begin{equation}
	\frac{\Gamma_{DA}}{\gamma_o} = \frac{6\pi\omega}{\epsilon_0c}\alpha_A''(\omega)|\mathbf{n}_A \cdot \bar{G}(\mathbf{r}_A,\mathbf{r}_D;\omega)\cdot \mathbf{n}_D|^2
	\label{FRET equation}
\end{equation}
where the polarizability is given by
\begin{equation}
	\alpha_A(\omega) = \lim_{\eta\rightarrow0} \frac{|\mathbf{p}_A|^2}{\hbar} \frac{1}{\omega - \omega_A + i\eta}
\end{equation}
in the rotating wave approximation. Note that this result gives rise to the same environmental dependence of FRET as compared to the semi-classical approach.

\textbf{Orientational averaging}. Here, we provide the generalized coordinate-free trace formulas for orientational averaging. The relevant quantity for the local density of states is given by
\begin{equation}
 	\braket{\mathbf{n}_D\cdot\bar{G}(r_D,r_D)\cdot \mathbf{n}_D}_D = \frac{1}{3}\,\text{Tr}[\bar{G}(r_D,r_D)].
 \end{equation} 
Similarly, one can show that the relevant quantity for the two-point spectral density is given by:
\begin{equation}
 	\braket{|\mathbf{n}_A\cdot\bar{G}(r_A,r_D)\cdot \mathbf{n}_D|^2 }_{D,A} = \frac{1}{9}\,\text{Tr}[\bar{G}^\dagger(r_A,r_D)\bar{G}(r_A,r_D)]
 \end{equation} 
where $\braket{\,}_{D,A}$ denotes orientational averaging over donor and acceptor orientations, and $^\dagger$ denotes the hermitian conjugate.

\textbf{Spherical nanoparticle system}. We provide details of the calculation for a spherical nanoparticle system. The formulation of the scattered dyadic Green function in spherically multilayered media was originally cast in \cite{li1994electromagnetic}. For self-consistency, we provide the results here in terms of a single summation rather than the double summation originally shown in \cite{li1994electromagnetic}. The simplification is achieved through the use of Legendre addition rules. 
\begin{equation}
	G_{rr} = \frac{ik_1}{4\pi}\sum_{n=1} n(n+1)(2n+1)\frac{h_{1a}h_{1d}}{\rho_{1a}\rho_{1d}} P_n(\cos\theta)R^f_p
	\label{Grr}
\end{equation}
\begin{equation}
	G_{r\theta} = \frac{ik_1}{4\pi}\sum_{n=1} (2n+1)\frac{h_{1a}h_{1d}'}{\rho_{1a}\rho_{1d}} P'_n(\cos\theta)\sin\theta\,R^f_p
	\label{Gro}
\end{equation}
\begin{equation}
	G_{\theta r} = \frac{ik_1}{4\pi}\sum_{n=1} (2n+1)\frac{h_{1a}'h_{1d}}{\rho_{1a}\rho_{1d}} P'_n(\cos\theta)\sin\theta\,R^f_p
	\label{Gor}
\end{equation}
\begin{eqnarray}
	G_{\theta\theta} = \frac{ik_1}{4\pi}\sum_{n=1} (2n+1) \left\{
	\frac{h_{1a}h_{1d}}{n(n+1)}P_n'(\cos\theta)\,R^f_s + \right.\nonumber \\ \left. 
	\frac{h'_{1a}h'_{1d}}{\rho_{1a}\rho_{1d}}\left(P_n(\cos\theta) - \frac{P'_n(\cos\theta)\cos\theta}{n(n+1)} \right)\,R^f_p \right\}
	\label{Goo} 
\end{eqnarray}
\begin{eqnarray}
	G_{\phi\phi} = \frac{ik_1}{4\pi}\sum_{n=1} (2n+1) \left\{
	\frac{h'_{1a}h'_{1d}}{\rho_{1a}\rho_{1d}}\frac{P'_n(\cos\theta)}{n(n+1)}\,R^f_p + \right. \nonumber \\ \left.
	h_{1a}h_{1d}\left(P_n(\cos\theta) - \frac{P'_n(\cos\theta)\cos\theta}{n(n+1)} \right)R^f_s \right\}
	\label{Goo} 
\end{eqnarray}
where $\theta=\theta_D-\theta_A$, $\rho_{kl}=k_k r_l$, $k_k = \sqrt{\epsilon_k}\omega/c$, and $G_{ij}=\mathbf{e}_{i_A}\bar{G}(r_A,r_D)\mathbf{e}_{j_D}$. Note that we have used the simplified notation $h_{kl}=h_n^{(1)}(\rho_{kl})$ and $h'_{kl}=\frac{d[\rho_{kl}h_n^{(1)}(\rho_{kl})]}{d\rho_{kl}}$ where $h_n^{(1)}(\rho)$ is the spherical hankel function of the first kind. The indices $i,j$ correspond the spherical coordinates $r,\theta,\phi$; the index $k=1,2$ corresponds to outer medium and inner medium with respect to sphere; and the index $l=d,a$ correspond to donor and acceptor respectively. $R_p^f$ and $R_s^f$ denote the centrifugal reflection coefficients for TM and TE polarized light as outlined in \cite{li1994electromagnetic}.

In the quasistatic regime, one can approximate these results to get an expression for the FRET FOM in terms of a dipole-driven multipolar source. For radial dipoles, the FRET figure of merit takes the form:

\begin{equation}
	F_{ET} = \left| 1 + \frac{r^3}{\kappa}\sum_{n=1}\frac{(n+1)^2\tilde{\alpha}_nP_n(\cos\theta_A) }{r_A^{n+2}r_D^{n+2}} \right|^2.
	\label{FET sphere}
\end{equation}
where $\tilde{\alpha}_n$ is the renormalized polarizability for the $n$th mode of the nanoparticle 
\begin{equation}
	\tilde{\alpha}_n = \alpha_n\left[1-i\frac{\alpha_n(n+1)k_1^{2n+1}}{n(2n-1)!!(2n+1)!!} \right]^{-1}
\end{equation}
and
\begin{equation}
	\alpha_n = \frac{n(\epsilon_2-\epsilon_1)}{n\epsilon_2 + (n+1)\epsilon_1}R^{(2n+1)}.
\end{equation}
The resonant frequency of the $n$th mode is $\omega_n=\sqrt{n}\omega_p/\sqrt{n\epsilon_\infty+(n+1)\epsilon_1}$. Similar expressions can be derived for other dipole orientations.

\section*{Acknowledgements}
We acknowledge funding from the National Science Foundation (NSF) (DMR-1654676).

\end{document}